\documentclass[manuscript]{aastex}

\def\arcsec{{$^{\prime\prime}$}}
\def\ptsec{$^{\prime\prime}\mskip-7.6mu.\,$}

\slugcomment{Version date: November 12, 2002}

\shortauthors{Bary et al.}
\shorttitle{H$_2$ Emission from T Tauri Stars}

\begin{document}

\def\plottwo#1#2{\centering \leavevmode
\epsfxsize=.5\textwidth \epsfbox[100 150 750 650]{#1} \hfil
\epsfxsize=.5\textwidth \epsfbox[100 150 750 650]{#2}}

\title{Detections of Ro-Vibrational H$_2$ Emission from the Disks of T~Tauri Stars}

\author{Jeffrey S. Bary\altaffilmark{1}, David A. 
        Weintraub\altaffilmark{1} and Joel H. Kastner\altaffilmark{2}}

\altaffiltext{1}{Department of Physics \& Astronomy,
Vanderbilt University, P.O. Box 1807 Station B, Nashville, TN 37235;
jeff.bary@vanderbilt.edu, david.a.weintraub@vanderbilt.edu}

\altaffiltext{2}{Carlson Center for Imaging Science, RIT, 
       54 Lomb Memorial Drive, Rochester, NY 14623; jhkpci@cis.rit.edu}


\begin{abstract}

We report the detection of quiescent H$_2$ emission in the v=1$\rightarrow$0~S(1) line at 2.12183~$\mu$m in
the circumstellar environment of two classical T~Tauri stars, GG~Tau A and LkCa~15, in high-resolution
(R~$\simeq$~60,000) spectra, bringing to four, including TW Hya and the weak-lined T~Tauri star DoAr~21, the
number of T Tauri stars showing such emission.  The equivalent widths of the H$_2$ emission line lie in the
range 0.02-0.10~\AA\ and, in each case, the central velocity of the emission line is centered at the star's
systemic velocity.  The line widths range from 9 to 14 km s$^{-1}$, in agreement with those expected
from gas in Keplerian orbits in circumstellar disks surrounding K-type stars at distances $\ge$~10~AU from the
sources.  UV fluorescence and X-ray heating are likely candidate mechanisms responsible for
producing the observed emission.  We present mass estimates from the measured line
fluxes and show that the estimated masses are consistent with those expected from the possible mechanisms
responsible for stimulating the observed emission.  The high temperatures and low densities required for
significant emission in the v=1$\rightarrow$0~S(1) line suggests that we have detected reservoirs of hot H$_2$
gas located in the low density, upper atmospheres of circumstellar disks of these stars.   

\end{abstract}

\section{Introduction}

The study of circumstellar disks around young stars may provide insight into how and when planets form.
Theoretical models suggest different mechanisms and timescales for planet formation.  The core accretion
model requires that a rocky core of $\sim$ 10M$_\oplus$ aggregate before substantial gaseous accretion can
form a Jupiter-like gas giant.  This theory predicts that gas giants will form a few AU or more from the star
due to the enormously greater mass of solids available beyond the snow line.  A time period of roughly
10$^7$ yr is predicted for giant planet formation to occur through runaway accretion in a greater than minimum
mass solar nebula \citep{liss1993,poll1996}.  A second though less thoroughly investigated model, which may
require a time period as brief as 10$^3$~yr for the formation of gas giant planets, proposes that such planets
may form as a consequence of gravitational instabilities in disks \citep{boss2000}.

Typically, astronomers have
resorted to emission from trace molecules and dust for insight into the physical conditions and masses of gas
in circumstellar disks.  Once these tracers disappear, the disks appear to have been removed from the system.
Two decades of such observations have shown that young stellar objects (YSOs) appear to be surrounded by thick
envelopes composed of dust and gas that evolve into circumstellar disks on the order of 10$^6$~yr
\cite{stro1995}.  The absence of thermal dust emission, as measured through millimeter and infrared continuum
excesses, as well as the non-detection of CO line emission towards most T~Tauri stars (TTS) over the ages of
3-5~Myr old, suggests that the disks become depleted in gas and dust grains rather quickly.  Two viable
explanations may account for these observations: in one model, when the disk becomes undetectable it no longer
exists, while in the second the disk is present but difficult to detect.  According to the first of the these
scenarios, circumstellar material has been dissipated through one or more of several processes:
photo-evaporation, accretion onto the star, tidal stripping by nearby stars, outflows, or strong stellar winds
\citep{holl2000}.  The second scenario, which describes the evolution of a circumstellar disk into a planetary
system according to the core accretion theory of planet formation, suggests that larger structures
(i.e., planetesimals) could have formed through accretion, collecting both solids and some gases like CO,
reducing the emitting surface area of the dust and pushing the emission from both dust grains and gaseous CO
below detectable levels.

In order to know whether the disks are dispersed or accretionally evolved, we must find a method to observe
the part of the disk that does not disappear quickly during accretion.  Since the bulk of the gaseous component
of the disk, hydrogen and helium gas, will be the last to be collected into protoplanets, assuming that it
avoids being removed through the various processes listed above, the H and He should remain in the
disk and in the gas phase even after the dust and gaseous CO has become undetectable.  Therefore, we need to
study the H$_2$ gas directly.

H$_2$, at the low temperatures typically found in circumstellar environments, predominantly populates
the ground vibrational level and only low rotational excitation levels and remains difficult to stimulate due
to its lack of a dipole moment.  In the small, inner regions of disks where temperatures may be
high enough to excite H$_2$ gas into the first vibrational state such that 2.12183 $\mu$m emission might be
possible, densities often are too high to permit detectable levels of H$_2$ line emission.  Therefore, the
bulk of the disk material has remained nearly impossible to detect directly.  Assuming that most of the gas in
these environments maintains a temperature much less than 1000 K, few molecules will be excited into the first
vibrational state; thus, the flux of thermally excited line emission from H$_2$ at 2.12183 $\mu$m will be
extremely low and consequently impossible to detect.  In the case of T Tau, the detected emission
from shocked H$_2$ traces approximately 10$^{-7}$~M$_\odot$ \citep{herb1996}, a mere fraction of the total
mass in the system \citep{wein1989a,wein1989b,wein1992}.

Pure rotational H$_2$ line emission found in the mid-infrared at 17 and 28 $\mu$m corresponds to excitation
temperatures on the order of $\sim$100 K.  These lines therefore should act as tracers of the bulk of the
unshocked gas which resides in the circumstellar disks.  Of these two lines, only the 17 $\mu$m line can be successfully
observed from ground based telescopes; however, Thi et al.\ (1999,2001a,b) reported the detection of emission
from H$_2$ at 17 and 28 $\mu$m using a low spatial and spectral resolution mid-infrared spectrometer on the
Infrared Space Observatory ({\it ISO}) towards GG~Tau, more evolved sources with Vega-type debris disks, and
Herbig Ae/Be stars.  However, in much more sensitive, higher spatial and spectral resolution mid-infrared
spectra obtained by Richter et al.\ (2002), centered on the 17 $\mu$m H$_2$ emission line, no emission was
observed towards GG~Tau, HD 163296, and AB Aur.  The Richter et al.\ results not only did not confirm the
detections made toward these stars, but also call into question other {\it ISO} detections.  

We have taken a different approach.  Motivated by models predicting that X-rays produced by TTS may be
sufficient to ionize a small fraction of the gas and indirectly stimulate observable near-infrared H$_2$
emission in circumstellar environments through collisions between H$_2$ molecules and nonthermal electrons 
\citep{gred1995,malo1996,tine1997}, we have begun a study of TTS searching for near-infrared H$_2$ emission.
We reported our initial detections of quiescent H$_2$ emission in the near-infrared towards TW Hya
\citep{wein2000}, an X-ray bright classical TTS (cTTS), and DoAr~21 \citep{bary2002}, an X-ray bright weak-lined
TTS (wTTS).  We previously have suggested that the most plausible explanation for the stimulation of the H$_2$
emission from these sources is the X-ray ionization process.  However, preliminary results from our modeling
of UV and X-ray photons originating from the source and incident on the disk at varying angles (Bary, in prep)
suggest that UV radiation from the source may be an equally likely stimulus for the observed emission.
Therefore, these results indicate that gaseous disks of H$_2$ surrounding other TTS possessing substantial
X-ray and UV fluxes could be detected at 2.12183 $\mu$m using high-resolution, near-infrared spectroscopy.

In this paper, we present the first results of a high-resolution, near-infrared spectroscopic survey of cTTS
and wTTS in the Taurus-Auriga and $\rho$~Ophiuchus star forming regions.  We report new detections of H$_2$
towards two sources: GG~Tau~A and LkCa~15.  We use the new data and our previously reported detections of
H$_2$ emission toward TW Hya and DoAr 21 to estimate disk masses for these four stars.

\section{Observations}

We obtained high-resolution (R~$\simeq$~60,000), near-infrared spectra of selected TTS (Table 1) in
Taurus-Auriga and $\rho$~Ophiuchus on 1999 December 26-29, and 2000 June 20-23, UT, using the Phoenix
spectrometer \citep{hink1998} on NOAO's 4-m telescope atop Kitt Peak.  In spectroscopic mode, Phoenix used a
256 $\times$ 1024 section of a 512 $\times$ 1024 Aladdin InSb detector array.  Our observations were made using
a 30\arcsec\ long, 0\ptsec8 (4 pixel) wide, north-south oriented slit, resulting in instrumental spatial and
velocity resolutions of 0\ptsec11 and 5~km~s$^{-1}$, respectively.  Our actual seeing limited spatial
resolution was $\sim$ 1\ptsec4.  The spectra were centered at 2.1218 $\mu$m, providing spectral coverage from
2.1167 to 2.1257~$\mu$m.  This spectral region includes three telluric OH lines, at 2.11766, 2.12325, and
2.12497~$\mu$m, that provide an absolute wavelength calibration.

Integration times for our program stars, with K magnitudes 5.9 $\leq$ K $\leq$ 9.6 (Table 1), ranged from
3600 s to 7200 s.  Nightly observations were obtained of an A0V star, either Vega or HD 2315, for telluric
calibration, with on-source integration times of 600 sec.  Flat-field images were
made using a tungsten filament lamp internal to Phoenix.  Observations were made by nodding the telescope
14\arcsec\ along the slit, producing image pairs that were then subtracted to remove the sky background and
dark current.   The spectra were extracted from columns of the array covering the region from 0\ptsec7 east to
0\ptsec7 west of each source for total beam widths of 1\ptsec4, beyond which no emission was detected.  Spectra
were then divided by the continuum to produce normalized spectra and ratioed with that of a star with a
featureless spectrum in this spectral region in order to remove telluric absorption features from the spectra
of the program stars.  SU Aur, a program star classified as an X-ray bright G2III wTTS produced a high signal
to noise ratio, featureless spectrum and therefore, also was utilized as a telluric calibrator.

\section{Results}

We detected line emission very nearly at 2.12183 $\mu$m from two cTTS, GG~Tau~A and LkCa~15 and, as reported
previously \cite{bary2002}, the wTTS DoAr~21 and the cTTS TW~Hya \cite{wein2000}\footnote{Note that LkCa~15 had
been classified as a wTTS (with an H$\alpha$ EW of 13 \AA) in the Herbig Bell Catalog; however, Wolk \& Walter
(1996) measured W$_\lambda$~(H$\alpha$)~=~21.9 \AA, leading some authors to reclassify this star as a
cTTS.}.  Full spectra for all our
target stars obtained with Phoenix are presented in Figure~1.  Most of the absorption features in these spectra
are telluric.  As best as could be done given tens-of-minutes timescale fluctuations in atmospheric conditions,
the spectra presented in Figure~2 have been corrected for telluric absorption features, extracted from the
total spectra presented in Fig.~1 and shifted in velocity using previously measured values of V$_{lsr}$
\citep{skru1993,kama2001}, placing the spectra in the rest frames of the stars.  In no cases in which H$_2$
emission was observed was any emission detected in the spectra beyond 0\ptsec7 from the central positions of
the star.  The central wavelengths of the emission lines were determined by fitting a Gaussian
to the emission features.  In each case, the central wavelength of the observed emission line matches the rest
wavelength of the 2.12183~$\mu$m~v=1$\rightarrow$0~S(1) transition of H$_2$ to within errors and the line is
narrow ($<$ 14 km s$^{-1}$) yet spectrally resolved ($>$~9~km~s$^{-1}$).  Therefore, we conclude that the
observed emission is from gaseous H$_2$ molecules within $\sim$~100 AU of the stars.  
The central velocities of the observed H$_2$ reservoirs are neither red-shifted nor
blue-shifted with respect to the stars and therefore experience no net line-of-sight motion.  H$_2$ equivalent
widths and line strengths for all four stars now detected in the H$_2$ line, corrected for extinction, are
presented in Table~1, along with 3-$\sigma$ upper limits for the line strengths of the non-detections.
A$_{\rm k}$ was determined using previously reported values for the visual extinction (Table~5) and taking
A$_{\rm k}$/A$_{\rm v}$ = 0.1 \cite{beck1978}.  

Several of our sources, notably V836 Tau, V819 Tau, and IP Tau show possible emission peaks at wavelengths just
shortward of 2.122 $\mu$m.  The low signal to noise levels of these stars' spectra, however, preclude drawing
any positive conclusions about these features in these data.  Two absorption features centered at
2.11699~$\mu$m and 2.12137~$\mu$m remain after the telluric corrections were made.  The line at 2.12137~$\mu$m
labeled `Al' is indicated in Figure 2.  The line at 2.11699 can be seen in many of the Fig. 1 spectra, in most
cases slightly red-shifted to $\gtrsim$~2.117~$\mu$m.  We have identified these lines as photospheric
absorption features produced by Al \citep{wall1996}.  Each of the stars with observed H$_2$ emission have both
of these photospheric features, as do many of the sources not detected in H$_2$ emission; however, four of our
sources do not.  The strength of these absorption features decreases toward earlier spectral types and
disappears entirely for the F and G type stars in our sample (Table 1) in agreement with the Wallace \& Hinkle
(1996) data.

\subsection{The Location of the Emitting H$_2$ Gas}

The spectrum of LkCa 15 is presented in Figure 2.   The double-peaked
H$_2$ emission line profile for LkCa~15 is consistent with the
double-peaked CO(J=2$\rightarrow$1) and HCO$^+$ emission profiles reported 
by Duvert et al.\ (2000).  The presence of blue-shifted and red-shifted
emission peaks for each of these emission features, including the
H$_2$, is indicative of gas revolving in a circumstellar disk
\citep{beck1993,mann1997}.  The peak-to-peak velocity separation for the
H$_2$ is resolved at $\Delta$v~$\approx$~10~$\pm$~1.5~km~s$^{-1}$, which is
considerably larger than the peak-to-peak separation of
$\Delta$v~$\approx$~2~km~s$^{-1}$ found from both the CO and HCO$^+$ spectra.
For a K5V star with a disk inclination angle of 34\degr $\pm$ 10\degr~as
determined by Duvert et al., the H$_2$ emission
would be produced at radial distances of between 10 and 30~AU while the
CO emission should arise from molecular gas located $\sim$$~$600~AU
from the source.  The combination of the CO and H$_2$ data indicate
that the near-infrared H$_2$ emission is sampling a different reservoir of gas than
is the CO(J=2$\rightarrow$1) emission and that the H$_2$ is from regions of the
circumstellar disk in which gas and ice giant planets might form.  In
addition, the mechanism responsible for exciting the H$_2$ apparently
is most efficient in doing so for gas at these intermediate distances
from LkCa~15.  As all four stars sharing H$_2$ emission have similar
H$_2$ line widths, fluxes and relative velocities, it is plausible
that the excitation mechanism is similar in all four stellar
environments.  We suggest, therefore, that the H$_2$ emission emerges
from the 10-30 AU region around each of these four stars.  Only
LkCa~15, however, has a disk inclined to our line of sight such that we
can distinguish the double-peaked profile (TW~Hya, in fact, is known to
be viewed nearly pole-on).  The spatial resolution for the spectra
centered on each source probes emission from the disks out to radii of
$\sim$~30~AU for TW~Hya and $\sim$~110~AU for LkCa~15, GG~Tau, and
DoAr~21.  Thus, the spatial resolution in our data is consistent with
our spectroscopically derived conclusion about the location of the emitting H$_2$.

In concluding that the gas is located in the regions 10-30 AU from the sources, we
are ruling out the possibility that the emission arises in extended halos
surrounding the sources.  The fact that in all four cases the H$_2$ line emission
is spatially unresolved is inconsistent with emission from a halo that should
be many hundreds or thousands of AU in extent.  In addition, since the H$_2$
emission lines appear to be spectrally resolved, the velocity line width of the
gas also is inconsistent with a halo interpretation for the observed emission, for
which we would expect line profiles with FWHM of only $\sim$ 1-2 km s$^{-1}$.  Line
widths of this size would be unresolved at R~$\simeq$~60,000.

\section{H$_2$ Mass Calculation}

While detection of H$_2$ is important in establishing the continued presence of gas in the evolving
circumstellar environment of TTS, a determination of the exact amount of H$_2$ gas still present has further
reaching implications concerning the timescale for planet formation and/or the dispersal of the disk.  Making
such a calculation requires understanding in what environments and under what conditions the H$_2$ molecule can
be excited into the {\it v} = 1, {\it J} = 3 state and what excitation mechanism(s) is active in this
environment. The v=1$\rightarrow$0~S(1) transition energy corresponds to a temperature of $\sim$~5000~K above
the ground state.  Therefore, excitation temperatures between 1000~K and 2000~K typically are required in all
types of astrophysical environments in order for H$_2$ gas to populate the first vibrationally excited level
and produce detectable levels of emission at 2.12183~$\mu$m \citep{tana1989}.

According to the disk model discussed in Glassgold et al.\ (2000), which includes X-ray heating, the {\it outer
surface layers} of the disk may be heated to temperatures of 1000~K out to several AU, while the midplane remains cool
at $\sim$~100~K.  Hollenbach et al.\ (2000) show that temperatures of 1000~K produced by incident UV radiation
are high enough to create photoevaporative flows that deplete disks on million year timescales.  Increasing the
temperature to 2000~K would enhance the photoevaporation process, destroy the disks on even shorter timescales,
and make detections of the emission we observed extremely rare or unlikely.  We note that both possible
stimulation mechanisms, UV fluorescence and X-ray ionization, are non-thermal processes.  Consequently, the
gas producing the emission may not be in Local Thermodynamic Equilibrium (LTE).

Proceeding under the assumption that the source is optically thin, we can determine the mass of H$_2$ emitting
v=1$\rightarrow$0~S(1) line emission from the measured line flux, independent of temperature, from

\begin{equation}
{\rm M(H_2)}_{\rm _{v=1\rightarrow0~S(1)}} = 1.76 \times 10^{-20} \frac{4 \pi F_{if} D^2}{E_{if} A_{if}},
\end{equation}

\noindent
where $F_{if}$ is the line flux measured in ergs s$^{-1}$ cm$^{-2}$ (column 9, Table~1), $E_{if}$ is the
energy and $A_{if}$ is the Einstein coefficient for the transition, $D$ is the distance in pc to the source,
and M(H$_2$) is given in units of M$_\odot$.  For the v=1$\rightarrow$0~S(1) transition,
$E_{10}$ = 9.338~$\times$~10$^{-13}$~ergs and $A_{10}$ = 2.09~$\times$~10$^{-7}$~s$^{-1}$ \citep{turn1977}.
The derived masses are given in column 2 of Table 2.

The next steps are to determine the population fraction $\chi_{v,J}(T)$, where {\it v} and {\it J} are the
vibrational and rotational number for the upper level of the transition, respectively, for H$_2$ in the
excited state located in the upper atmosphere of the disk in the 10-30 AU emission volume, and the mass
fraction of the disk represented by this fractional part of the disk.  One approach is to assume LTE in which
case the total mass in the emitting volume of the disk would be

\begin{equation}
\frac{{\rm M(H_2)}_{\rm _{v=1\rightarrow0~S(1)}}}{\chi_{v,J}(T)}.
\end{equation}

\noindent
Under LTE conditions at 1500~K, $\chi_{v,J}(T)$~=~5.44~$\times$~10$^{-3}$ and the masses range from
6.4$\times$10$^{-10}$~M$_\odot$ for TW Hya to 8.1$\times$10$^{-8}$~M$_\odot$ for DoAr~21 (column 3, Table~2).
If the gas is not in LTE or is at a lower temperature, then $\chi_{v,J}$ would be smaller and the masses
larger.  As is evident from these calculations, the masses estimated from `hot' H$_2$ are only small fractions
of the total disk masses, based on comparisons with disk gas masses estimated from observations of warm H$_2$,
cool dust, and cold CO (see Table~2).  Since the 2.12183 $\mu$m observations of hot H$_2$ do not likely sample
gas present in the inner disk (r$\lesssim$10~AU), the dense midplane of the disk at 10-30 AU, nor the disk
outside of 30~AU, what is needed is a (temperature independent) scaling factor, {\it f}, such that

\begin{equation}
{\rm M}_{disk} = f {\rm M(H_2)}_{\rm _{v=1\rightarrow0~S(1)}} .
\end{equation}

\noindent
Just as canonical scaling factors have been determined for the mass ratio between dust and gas and for the
number ratio of CO and H$_2$, we have calculated {\it f} from equation (3) and the ``line emission'' masses in
Table~2 for our four stars to determine whether such a scale factor relating hot H$_2$ to total disk mass can
be derived by using previous estimates of M$_{disk}$ in Table~2.  Scale factors derived from masses implied by
several different tracers of the total disk mass are listed in Table~3.  The scale factors all fall within the
range from $\sim$10$^{7}$ to $\sim$10$^{9}$ and suggest that we are detecting as little as one billionth of
the total disk mass.

The differences in these mass estimates, and therefore the range in {\it f} values, most likely arise because
the various observations are sensitive to different reservoirs of material, with the (sub)mm dust continuum
consistently giving the largest {\it f} values and the CO and warm H$_2$ smaller {\it f} values.  The emission
from H$_2$ at 2.12183~$\mu$m must arise from hot gas, at T~$>$~1000~K, which necessarily must reside near the
surface of the disk and within a couple of tens of AU of the star (see \S 3.1 and \S 5), while the millimeter
continuum and CO observations trace cooler gas at distances of 100s of AU.  Given the location and density of
the hot gas in comparison to the cool gas, it is reasonable to conclude that the tracers of the cooler gas will
uncover much more mass than the hot gas tracer.

If the hot gas is stimulated by non-thermal processes such as X-ray ionization and/or UV fluorescence, as we
argue is likely in \S 5, the near-infrared H$_2$ line emission will be most sensitive to gas in the inner few
tens of AU of the disk.  In the case of DoAr 21 for which previous non-detections of cold dust indicated that
this star was largely void of circumstellar material, our detection of quiescent H$_2$ emission at
2.12183 $\mu$m suggests that a gaseous component in the inner few tens of AU of the disk persists even after
the cold gas tracers stop producing detectable levels of emission \citep{bary2002}.  Therefore, we conclude
that high-resolution, near-infrared spectroscopy of H$_2$ provides a means to detect evolved, low mass disks
which may have no other observable signatures at this point in their evolution.

\section{Identifying the Stimulation Mechanism}

H$_2$ emission in the near-infrared has been observed in a variety of stellar environments such as planetary
nebulae, Herbig-Haro objects, and galaxies.  For example, H$_2$ emission observed from planetary nebulae and
HH objects is attributed to heating of molecular gas as it encounters shock fronts from outflows associated
with the source.  UV radiation produced in accretion flows or emitted by the sub-dwarf embedded in a PN can
likewise stimulate H$_2$ molecules located near these sources to fluoresce.  Near-infrared H$_2$ emission also
has been observed towards a Seyfert galaxy, with X-rays suggested as the mechanism responsible for the
excitation of the molecules \citep{tine1997}.  Similarly, several authors have proposed that gas densities and
X-ray luminosities in the environments of TTS also are sufficient to produce H$_2$ emission in the
near-infrared, with the strongest transition being the v=1$\rightarrow$0 S(1) ro-vibration line
\citep{gred1995,malo1996,tine1997}.

With the high-resolution velocity information from our data placing the velocity of the gas at the systemic
velocities of the stars with only a very small dispersion, and the spatial resolution of and spectral
information from our data placing the H$_2$ emission within a few tens of AU of the respective stars, we conclude
that the observed emission is unlikely to originate from shock excitation in a YSO outflow, for which
typical outflow velocities are at least many tens of km~s$^{-1}$ and the shocked gas is most often observed at
great distances from the stars.  The FWHM velocity distribution for the observed emission lines, 
$\sim$~9 to 14~km~s$^{-1}$, is larger than the thermal broadening expected from H$_2$, but not large enough to
be produced by gas shocked by a jet, even for a jet aligned perpendicular to the line of sight with a small
opening angle.  For example, assuming an extremely small opening angle of $\sim$~4\degr, as has been observed
for the outflow observed towards RW~Aur \citep{doug2000}, and a launch velocity of  $\sim$~200~km~s$^{-1}$, the
H$_2$ line would be broadened by $\simeq$~10~km~s$^{-1}$.  Even a tiny inclination angle of the jet toward our
line of sight would increase the breadth of the line tremendously by creating blue- and red-shifted jets of
gas.  Since all of our sources' emission lines have similarly small FWHM, it is statistically reasonable to
argue that most of these sources do not possess jets aligned nearly perpendicular to the line of sight,
although conceivably one of the sources might be so aligned.  Therefore, we conclude that shock excitation in
a stellar outflow is not likely to be the stimulating mechanism for most and probably not any of these sources.

Since the observed line widths appear to rule out shocks in outflowing gas, the velocity dispersion in our line must be due only
to thermal and/or orbital broadening, leaving ultraviolet pumping and X-ray ionization as candidate mechanisms
for stimulating the H$_2$ emission.  Ultraviolet pumping and X-ray ionization would serve to enhance the
population of H$_2$ molecules in excited states in the circumstellar disks of young stars via photon
excitation (UV) or secondary interactions between H$_2$ molecules and ions produced by X-ray ionization
events.  Using the disk model found in Glassgold et al.\ (2000) and cross-sections for H$_2$ molecules as
``seen'' by UV photons and X-rays determined from Yan et al.\ (1998) we have calculated a set of optical depth
surfaces in the disk, in the range 0.2~$\le$~$\tau$~$\le$~2, to which photons of different energies and angles
incident upon the inner disk would penetrate.  The optical depth surfaces determined by our model assume a
disk composed entirely of H$_2$ molecules with incident photons originating at an inner radius of 0.35 AU.
Although this model is simple, the results from our preliminary simulations (Bary, in prep) suggest that both
UV and X-ray photons could be responsible for the observed 2.1218~$\mu$m emission.  Because the interaction
cross-sections for the X-rays are much smaller than those for UV photons, X-rays will penetrate through a
greater optical column than will UV photons.  Thus, some X-ray photons will penetrate deeper below the surface
layer of the disk than will the UV photons and deposit their energy in regions of the disk in which the number
densities are greater than 10$^7$~cm$^{-3}$; such densities are too high to produce the observed H$_2$
emission at radii less than 50 AU.  At these densities, excited H$_2$ molecules will be
collisionally de-excited.  However, a significant fraction of the X-ray photons will be absorbed at the more
shallow disk layers where the number density is below 10$^7$~cm$^{-3}$.  The fraction of X-ray photons absorbed in regions
of greater vs. smaller number density will depend on the exact physical shape (flaring profile) of the disk,
the vertical density gradient in the disk, and the place of origin of the X-ray photons (stellar surface or
accretion funnels). A majority of UV photons incident upon the surface of the disk at distances between 10 and
30 AU will be absorbed at distances of 5-10 AU above the mid-plane, in a region of the disk where the number
densities (10$^4$~$\le$~$n$~$\le$~10$^7$~cm$^{-3}$) are suitable for producing observable H$_2$ emission at
2.12183~$\mu$m.

Because both UV and X-ray photons are capable of penetrating to and being absorbed at depths
and radii at which the observed line emission likely arises, in the following discussion, we examine both UV
fluorescence and X-ray excitation models to determine if the flux of emission observed towards each
source can be stimulated by either or both mechanisms.


\subsection{X-ray Stimulated Emission}

We now predict the flux of the v=1$\rightarrow$0 S(1) line emission using the models of X-ray
excitation presented by Maloney et al.\ (1996).  For example, using the ROSAT measured X-ray flux from
DoAr 21, we determined the X-ray energy deposition rate per volume element H$_x$/n to be
3-8~$\times$~10$^{-27}$~ergs~s$^{-1}$~cm$^{-3}$ at a distance of 10 AU from the source in a disk and assuming
a number density\footnote{Note that this density is below the critical density and all of
the H$_2$ line flux predicted to be stimulated by the X-rays will be observable.  However, even if far less
than 50\% of the X-rays are absorbed in a less than critically dense region and the predicted line fluxes are
cut in half, the X-ray mechanism still proves to be effective at producing most of the observed H$_2$ emission
from these sources.} of 10$^5$~cm$^{-3}$.  For this range in likely values of
H$_x$/n, we can estimate the emergent integrated surface brightness in the 1$\rightarrow$0 S(1) emission line
from Figure 6a of Maloney et al.\ (1996).  Assuming the emission emerges from an annulus in a circumstellar
disk stretching from 10 to 30 AU and at a distance of 160 pc, the predicted flux of the
1$\rightarrow$0 S(1) line lies in the range 2.4~$\times$~10$^{-17}$ to
2.4~$\times$~10$^{-14}$~ergs~s$^{-1}$~cm$^{-2}$ (Table~4).  The observed line flux falls within and near the
top of this predicted range.  Thus, this calculation suggests that the observed emission could be attributed
solely to X-ray excitation.  Similar results (Table~4) are found using the X-ray fluxes for GG Tau, TW Hya,
and using an assumed X-ray flux for LkCa~15.

Thus, for the stars that are sources of X-rays, we have determined that X-rays are capable of stimulating the
observed H$_2$ emission at 2.12183~$\mu$m.  In fact, two of the four sources for which we have detected the
quiescent H$_2$ emission, DoAr~21 and TW~Hya, are among the X-ray brightest TTS (column 6, Table~1).
However, GG~Tau is considerably less X-ray luminous, while LkCa~15 was not detected at all by ROSAT.  If
X-rays are important for stimulating the observed emission, we must understand why these four stars appear so
different in observed X-ray emission.

Historically, we know that most wTTS are strong X-ray emitters, with many wTTS having been discovered through
extensive X-ray surveys of star forming regions (e.g., Casanova et al.\ 1995; Neuh\"{a}user et al.\ 1995).  The
apparent bimodal distribution of TTS as X-ray sources, in which wTTS were almost always detected and cTTS were
much less often detected, led to widespread acceptance that the X-ray emission mechanism was related to the
evolution of the star/disk system.  According to this scenario, X-ray emission from wTTS is attributed to
conservation of angular momentum and ``spinning up'' of the star as the magnetic field lines decouple
from the dissipating circumstellar disk, allowing the forming star to contract \citep{edwa1993,bouv1993}.
With the decrease in rotational period of the source, surface magnetic activity presumably increases.  In turn,
this process is expected to lead to an increase in coronal heating and the production of detectable levels of
X-ray emission.  Thus for several years, X-ray emission was thought to distinguish wTTS from cTTS, with
observational evidence suggesting that wTTS were, indeed, fast rotators \citep{choi1996,shev1998} while cTTS
rotated more slowly due to disk-locking and therefore lacked the mechanism for X-ray production.  With
increasing statistics from large X-ray surveys, however, many cTTS also have been found to be generating
X-rays.  Since X-rays are no longer thought to distinguish cTTS from wTTS \citep{feig1999}, it is no longer
clear what mechanism produces X-rays in the YSO environment.  In addition, increasing evidence has called into
question the existence of a bimodal rotation period distribution for wTTS and cTTS \citep{stas1999,rebu2001},
with some authors now suggesting such a distribution applies only to TTS in the mass range of about
0.25-1.0~M$_\odot$ \cite{herb2002}. 

In recent work based on Chandra spectra, Kastner et al.\ (2002) use the inferred plasma temperature distribution
and the relative densities of iron, oxygen and neon deduced from high-resolution Chandra X-ray spectroscopy to
infer that the bulk of the X-ray emission from TW Hya is
generated via mass accretion from its circumstellar disk.  If this is correct, and more so if this is a
common mechanism for X-ray production, then X-ray bright TTS --- including wTTS --- would be those still
accreting material from their disks.  This interpretation is contrary to the common wisdom suggesting X-rays
are dominantly from fast rotating wTTS that are no longer accreting material from circumstellar disks and
suggests that DoAr~21, despite being a wTTS, not only has H$_2$ in a circumstellar disk, but may still be
actively accreting material from a circumstellar disk \citep{bary2002}.

We suggest that despite the absence of detectable levels of X-ray emission, LkCa~15 could be accreting from a
circumstellar disk; however, because its thick disk is oriented nearly along our line of sight, the disk may
absorb a significant fraction of the X-ray emission, thereby decreasing the X-ray flux that escapes the
circumstellar region to a level that is below detection thresholds.  A column of 10$^{21}$ cm$^{-3}$
(equivalent to A$_{\rm v}$~$\sim$~0.5; A$_{\rm v}$~$\sim$~0.64 for LkCa~15 (see Table~5)) is sufficient to
decrease the flux from a source with an intrinsic X-ray luminosity of L$_x$ = 
3~$\times$~10$^{29}$~ergs~s$^{-1}$ to a count rate below 0.003~s$^{-1}$, which is well below the
detection threshold of the ROSAT all sky survey.  A scenario such as this may explain the observed differences
in X-ray fluxes from cTTS and wTTS, since most cTTS are observed to have thicker disks more capable of
attenuating the flux of X-rays produced within the inner boundary of the disks.

A star that previously has stood out as a notable exception to the absence of X-ray bright cTTS is TW Hya.
TW Hya has been imaged extensively at wavelengths ranging from the optical \cite{kris2000} and near-infrared
\cite{wein2002} through the mm \cite{wiln2000}, and all of these observations indicate the presence of a
face-on disk.  Therefore, in the case of TW Hya, there is no line-of-sight gas and dust associated with the
circumstellar disk of the star to attenuate the X-ray emission from the star.

\subsection{Ultraviolet Stimulated Emission}

In an attempt to determine whether that the UV flux (F$_{\rm UV}$) is capable of
producing the observed H$_2$ line flux, we compare observed and extrapolated values of F$_{\rm UV}$ with
those predicted from the models of Black and van Dishoeck (1987).  Using the model for UV fluorescence found
in Black and van Disheock (1987), the predicted UV flux is given by

\begin{equation}
{\rm F_{\rm UV}} = \frac{4\pi {\rm F_{\rm v=1\rightarrow0~S(1)}}}{{\Omega_{\rm Disk}} \varepsilon {\it f}_{\rm 1\rightarrow0}}, 
\end{equation}

\noindent
where $\varepsilon$ is the UV absorption efficiency, {\it f}$_{\rm 1\rightarrow0}$ is the fraction of the
total infrared flux emitted in the the 2.12183~$\mu$m emission line, ${\Omega_{\rm Disk}}$ is
the solid angle of the disk that intercepts the stellar UV flux and F$_{\rm v=1\rightarrow0~S(1)}$ is
the observed H$_2$ line flux in erg~s$^{-1}$~cm$^{-2}$.  Using the scale height of the disk at a distance of
30 AU from the source, we conservatively estimate $\frac{{\Omega_{\rm Disk}}}{4\pi}$ to be $\approx$~0.1 since
UV photons are likely to be absorbed at distances above the midplane that are larger than the scale height.
The model dependent UV absorption efficiencies and line fractions  were chosen from models 12 and 15 from
Black and van Dishoeck (1987) to provide a wide range of predicted UV fluxes.

Of the four sources, TW~Hya is the only
star that has been observed in the wavelength range, 925-1130 \AA, which are the wavelengths of interest
because only photons with $\lambda$~$\leq$~1130~\AA~possess the right energy to produce infrared H$_2$
fluorescent emission.  Far Ultraviolet Space Explorer {\it FUSE} observations (Herczeg, private communication)
show that the uncorrected line and continuum fluxes for TW~Hya have nearly the same magnitude when integrated
over the wavelengths 925-1130 \AA.  Using equation (4), we predict a range of
6.6~$\times$~10$^{-12}$~$\leq$~F$_{\rm UV}$~$\leq$~4.8~$\times$~10$^{-10}$~erg~s$^{-1}$~cm$^{-2}$ for the
value of UV flux from TW~Hya necessary to stimulate the observed H$_2$ line emission.  Using the
sum of the integrated {\it FUSE} continuum flux and the line fluxes (CIII at 977 \AA, O VIII at 1032 and
1038~\AA) for TW~Hya we estimate that the model dependent UV flux is responsible for 1\% to 20\% of the
observed H$_2$ line emission at 2.12183 $\mu$m.

For the other sources, we converted the {\it U} magnitude for each star to a flux at 3620 \AA.  Assuming the
stars radiate as blackbodies with temperatures indicated by their main-sequence spectral types, we
extrapolated the 3620 \AA~flux to 1000 \AA.  By comparing the theoretical F$_{\rm UV}$ for TW~Hya to the
observed value of F$_{\rm UV}$, we can determine if the extrapolation method accurately predicts the UV
continuum flux.  For TW~Hya, we determined F$_{\rm UV}$ to be about 3~$\times$~10$^{-18}$~erg~s$^{-1}$~cm$^{-2}$, about
five orders of magnitude less than the level observed by {\it FUSE}.  We conclude that the extrapolation method
for determining F$_{\rm UV}$ is a poor predictor of the stellar UV flux.  Therefore, for DoAr~21, LkCa~15, and
GG~Tau, without {\it FUSE} observations we can only surmise that their F$_{\rm UV}$ spectra may be similar to
that of TW~Hya and, thus, that their actual levels of F$_{\rm UV}$ may be capable of stimulating at least part
of the observed 2.12183~$\mu$m line emission.

The results of the UV fluorescence model suggest that a significant fraction of the infrared H$_2$ emission
may be produced via UV photo-excitation for each star.  However, the model is not well
constrained.  For example, UV emission can be absorbed by circumstellar and interstellar gas, in which case,
the UV line and continuum fluxes would be lower limits.  Predicting the UV flux necessary to produce the
2.12183 $\mu$m line flux is complicated by the narrow range of densities for which the efficiencies and line
fractions have been calculated.  In addition, Black and van Dishoeck (1987) assume a plane parallel cloud of
gas with a constant density; in our case we used 3~$\times$~10$^{3}$~cm$^{-3}$, while in a realistic
circumstellar disk there will be a density gradient with the outer layers of the disk having lower densities
and the inner layers of the disk having much higher densities.  As shown in Black and van Dishoeck~(1987), the
efficiencies can be affected greatly by the column and space densities of the gas.

Ultraviolet excesses associated with pre-main-sequence stars normally are attributed to active accretion of
disk material onto a forming star.  Thus, if X-rays are produced in accretion flows one would expect to
find that those same sources show ultraviolet excesses.  We have calculated {\it U$-$V} color excesses towards
the H$_2$ detected sources (Table~5) by dereddening their {\it U} and {\it V} magnitudes and comparing the
observed, dereddened {\it U$-$V} colors with standard main-sequence colors \citep{john1966}.  Using the
photometry of GG Tau Aa and Ab from Ghez et al.\ (1997), we determined a value for the excess of each
component of the northern binary.  TW~Hya, DoAr~21, and GG~Tau~Aa and GG~Tau~Ab all have significant excesses
with [{\it U}$-${\it V}]$_{excess}$~$<$~$-$1.3.  LkCa 15 appears to have a smaller, but significant,
ultraviolet excess.  On this basis we conclude that each of these sources may still be undergoing active
accretion and producing excess UV emission that could be responsible for stimulating some of the observed
H$_2$ emission.

A closer look at the ultraviolet excesses from TW~Hya and LkCa~15 suggests a possible dependence of
[{\it U}$-${\it V}]$_{excess}$ on the inclination angle of the disk.  The nearly edge-on disk of LkCa~15 may
absorb a significant fraction of the ultraviolet emission produced in the accretion process, just as suggested
for the attenuation of its X-ray flux.  On the other hand, TW~Hya, which has been imaged and shown to be viewed
nearly pole-on, produces the largest value of [{\it U}$-${\it V}]$_{excess}$ of any of these four sources.
Based on the shape of the H$_2$ emission line from DoAr~21, we inferred \cite{bary2002} the orientation of its
disk to be more pole-on ({\it i}~$>$~55\degr) than that of LkCa~15; such an orientation is consistent with the
larger value of [{\it U}$-${\it V}]$_{excess}$ found for DoAr~21 than for LkCa~15.  Due to the complexity of
the GG~Tau system, with both circumbinary and circumstellar material, we are unable to relate the
[{\it U}$-${\it V}]$_{excess}$ values for these stars to disk orientation in any meaningful manner.  While our
sample remains small so that we lack the statistical certainty that could prove the existence of a correlation
between the strength of [{\it U}$-${\it V}]$_{excess}$ and disk orientation, such a possible correlation is
intriguing and consistent with the data presented.

\section{Color-Color Diagrams}

Near-infrared broad band photometry of YSOs previously has been used to identify sources that are likely
candidates for harboring circumstellar disks. In this section, we examine infrared color-color diagrams to 
determine if there exists a correlation between the infrared excess, the implied presence of a 
circumstellar disk and the observed H$_2$ emission. {\it J$-$H}~vs.\ {\it H$-$K} color-color diagrams of YSOs can
be used to distinguish sources possessing infrared excesses from main-sequence stars.  However, as described by
Haisch et al.\ (2000), {\it JHK} observations do not sample long enough wavelengths to allow for unambiguous
identification of sources that posses circumstellar disks.  Lada et al.\ (2000) explain that the {\it JHK}
near-infrared excess depends on the star/disk system (i.e., disk inclination, accretion rate, presence and
size of holes in the inner disk) and also may be produced by emission from HII regions and reflection nebulae
usually associated with star forming regions.  {\it L} band observations, however, are much less sensitive
to emission from HII regions and scattered light from reflection nebulae and will measure infrared excess
emission from circumstellar disks independent of the star/disk system.  Therefore, an infrared excess observed
using {\it JHKL} photometry can be characterized more confidently as being produced by a circumstellar disk.
To this end, extensive {\it L} band surveys are being conducted to determine disk fractions in many nearby
young star clusters \citep{lada2000,hais2000,hais2001a,hais2001b,keny2001}.

Using {\it JHKL} observations made by Rydgren et al.\ (1976), Rucinski \& Krautter (1983), Kenyon \& Hartmann
(1995), Strom et al.\ (1995), Coulson et al.\ (1998), and Calvet et al.\ (2002), we have plotted
the TTS in our survey on {\it JHK} (Figure~3) and {\it JHKL} (Figure~4) color-color diagrams.
The solid main-sequence lines included on the diagrams are drawn from standard colors found in Koorneef
(1983).  With the exception of LkCa~15, none of the sources display unambiguous evidence of an infrared excess
based solely upon the {\it JHK} observations.  However, the {\it JHKL} color-color diagram shows clear
evidence of infrared excesses associated with GG~Tau~Aa, GG~Tau~Ab, LkCa~15, and TW~Hya.  All four of these
have been observed to have other evidence of circumstellar material, either millimeter continuum observations
and/or CO line emission.  DoAr~21 shows only a marginal hint of a {\it K$-$L} excess, as the error bar of its
{\it K$-$L} color overlaps with the interstellar reddening vector for a K1V star.  Therefore, we conclude 
that DoAr~21 lacks strong evidence of a near-infrared excess, in good agreement with previous
non-detections of millimeter continuum observations of this source \citep{andr1990,andr1994}.  More accurate
{\it K} and {\it L} band photometry of DoAr~21 will determine its true position on the {\it JHKL} diagram and
provide a more definitive conclusion as to the presence or absence of an infrared excess.

\subsection{A Variety of Circumstellar H$_2$ Detections}

Several detections of H$_2$ in the vicinity of young stellar objects have been made in recent years.  As
previously mentioned, {\it ISO} observations of Herbig Ae/Be stars and cTTS claim to have detected pure rotational
emission from H$_2$ in the vicinity of these stars \cite{thi2001b}.  Ratios of the mid-infrared lines give
temperatures of the emitting gas to be approximately $\sim$~100~K placing the gas in the outer regions of the
disk at radii beyond 100 AU.  These observations have yet to be confirmed and Richter et al.\ (2002) have
called these interpretations into question.  Herczeg et al.\ (2002) have
detected far ultraviolet H$_2$ emission from the cTTS, TW Hya.  They interpret this emission as
produced by Lyman $\alpha$ pumping of H$_2$ molecules located within 0\ptsec05, or $\sim$~3~AU, of the source.
In contrast to the pure rotational emission which may have been observed by {\it ISO}, the detections reported
by Herczeg et al.\ (2002), Weintraub et al.\ (2000), Bary et al.\ (2002) and herein, represent the only
detections of H$_2$ in the inner regions of TTS disks where planet formation may occur.

\section{Conclusions}

We have demonstrated that high-resolution, near-infrared spectroscopy of young stars can provide useful
information about circumstellar gas around TTS.  The discovery of ro-vibrational emission from quiescent H$_2$
towards GG~Tau A and LkCa~15 adds to the increasing number of such observations \citep{wein2000,bary2002}.
These data suggest that the gaseous component of circumstellar disks persists and may be substantial, even
though other tracers of the gas may suggest otherwise.  However, limited by the spatial resolution of our data,
we have inferred the proximity of the emitting H$_2$ gas to the source from kinematic arguments, while
further observations such as high-resolution narrow band images are needed to definitively identify the
location of the H$_2$ emission.  Spectral observations of other ro-vibrational H$_2$ emission lines are needed
in order to determine a gas temperature and help confirm the excitation mechanism for the gas.  Observations
such as these will lead to a more complete understanding of the emission we have detected and the implications 
thereof to planet formation.

\noindent
\begin{deluxetable}{lllllcccrc}
\tabletypesize{\scriptsize}
\tablecolumns{10}
\tablenum{1}
\tablewidth{0pt}
\tablecaption{WTTS and CTTS: H$_2$ Detections and Non-detections}
\label{tbl-1}
\tablehead{
		 \colhead{Source}
	       & \colhead{Class}
	       & \colhead{Sp.\tablenotemark{a}}
	       & \colhead{K}
	       & \colhead{H$\alpha$\tablenotemark{b}}
	       & \colhead{log L$_x$\tablenotemark{c}}
	       & \colhead{Al}
	       & \colhead{Al}
	       & \colhead{H$_2$ Line}
	       & \colhead{H$_2$}  \\
	         \colhead{}
	       & \colhead{}
	       & \colhead{Type}
	       & \colhead{(mag)}
	       & \colhead{EW}
	       & \colhead{(ergs s$^{-1}$)}
	       & \colhead{(2.117}
	       & \colhead{(2.121}
	       & \colhead{Flux}
	       & \colhead{EW} \\
	         \colhead{}
	       & \colhead{}
	       & \colhead{}
	       & \colhead{}
	       & \colhead{(\AA)}
	       & \colhead{}
	       & \colhead{$\mu$m)}
	       & \colhead{$\mu$m)}
	       & \colhead{(erg s$^{-1}$ cm$^{-2}$)}
	       & \colhead{(\AA)}
	  }      
\startdata

IP Tau   & cTTS  & M0V    		   & 8.44 		  & 12.7   & 29.5             &yes & yes  & $<$ 1.2$\times$10$^{-15}$ &\nodata     \\
IQ Tau   & cTTS  & M0.5V  		   & 8.16 		  &  7.7   & \nodata          &yes & yes  & $<$ 3.5$\times$10$^{-15}$ &\nodata     \\
GG Tau A & cTTS  & K7/M0.5V   		   & 7.3  		  & 40     & 29.4             &yes & yes  &     6.9$\pm$0.5$\times$10$^{-15}$ \tablenotemark{d} & 0.10       \\
V819 Tau & wTTS  & K7V    		   & 7.97 		  &  3.2   & 30.2             &yes & yes  & $<$ 3.0$\times$10$^{-15}$ &\nodata     \\
V836 Tau & wTTS  & K7V    		   & 8.75 		  &  4.4   & 29.8             &yes & yes  & $<$ 9.4$\times$10$^{-16}$ &\nodata     \\
GSS 29   & wTTS  & K7-MO\tablenotemark{e}  & 8.19		  &\nodata & 30.4             &yes & yes  & $<$ 1.6$\times$10$^{-15}$ &\nodata     \\
TW Hya   & cTTS  & K7V    		   & 7.37		  & 220    & 30.3             &\nodata& yes &   1.0$\pm$0.1$\times$10$^{-15}$ \tablenotemark{d} & 0.02     \\
LkCa 15  & cTTS  & K5V    		   & 7.6  		  & 13     & \nodata          &yes & yes  &     1.7$\pm$0.2$\times$10$^{-15}$ \tablenotemark{d} & 0.05       \\
RXJ0516.3& wTTS  & K4V    		   & 9.65\tablenotemark{f}&  0.1   & 30.6             &yes & \nodata\tablenotemark{g}   & $<$ 1.8$\times$10$^{-15}$ &\nodata     \\
SR 12    & wTTS  & K4-M2.5\tablenotemark{e}& 8.1		  &  4.5   & 30.4             &yes & yes  & $<$ 1.9$\times$10$^{-15}$ &\nodata     \\
DoAr 25  & wTTS  & K3-M0\tablenotemark{e}  & 7.57		  &  2.3   & 29.4             &yes & yes  & $<$ 2.2$\times$10$^{-15}$ &\nodata     \\
DoAr 21  & wTTS  & K1V    		   & 6.13		  &  0.8   & 31.2             &yes & yes  &     1.5$\pm$0.9$\times$10$^{-14}$ \tablenotemark{d} & 0.06       \\
HD 283572& wTTS  & G5IV   		   & 6.53 		  &  1.6   & 31.2             &no  & no   & $<$ 5.2$\times$10$^{-15}$ &\nodata     \\
SU Aur   & wTTS  & G2III  		   & 5.86 		  &  3.5   & 30.5             &no  & no   & $<$ 6.1$\times$10$^{-15}$ &\nodata     \\
HD 34700 & wTTS  & G0V    		   & 7.45\tablenotemark{f}&\nodata & \nodata          &no  & no   & $<$ 2.6$\times$10$^{-15}$ &\nodata     \\
S1/GSS 35& wTTS  & F2II   		   & 6.29		  &  3.5   & 30.1             &no  & no   & $<$ 3.8$\times$10$^{-15}$ &\nodata     \\

\enddata
\scriptsize
\tablenotetext{a}{ Kenyon et al.\ (1998), N\"{u}rnberger et al.\ (1998), Wolk \& Walter (1996), Hamann \& Persson (1992), Chen et al.\ (1995), Leinert et al.\ (1993)}
\tablenotetext{b}{ Herbig \& Bell (1988), Kenyon et al.\ (1998), Bouvier \& Appenzeller (1992), Webb et al.\ (1999), Allencar \& Basri (2000)}
\tablenotetext{c}{ Neuh\"{a}user et al.\ (1995), Casanova et al.\ (1995), Kamata et al.\ (1997), Magazz\'{u} et al.\ (1997)}
\tablenotetext{d}{ The H$_2$ line flux value is extinction corrected.  For stars with more than one value for A$_{\rm v}$ (Table 6), the larger A$_{\rm v}$ was used to obtain an upper limit.}
\tablenotetext{e}{ Bouvier \& Appenzeller (1992) list spectral types as peculiar and give two possible spectral types.}
\tablenotetext{f}{ Estimated}
\tablenotetext{g}{ Cannot be determined due to poor signal to noise in spectrum.}
\end{deluxetable}

\begin{deluxetable}{llcccc}
\tabletypesize{\scriptsize}
\tablecolumns{6}
\tablenum{2}
\footnotesize  
\tablewidth{0pt}
\tablecaption{Gas Mass Estimates}
\label{tbl-2}
\tablehead{
		 \colhead{Source}
	       & \colhead{hot H$_2$\tablenotemark{a}}
	       & \colhead{hot H$_2$\tablenotemark{b}}
	       & \colhead{warm H$_2$}
	       & \colhead{CO emission}
	       & \colhead{(sub)mm cont.} \\
	         \colhead{}
	       & \colhead{(line emission)}
	       & \colhead{(LTE)}
	       & \colhead{(17 \& 28 $\mu$m)}
	       & \colhead{}
	       & \colhead{} \\
	         \colhead{}
	       & \colhead{(M$_\odot$)}
	       & \colhead{(M$_\odot$)}
	       & \colhead{(M$_\odot$)}
	       & \colhead{(M$_\odot$)}
	       & \colhead{(M$_\odot$)} \\ 
 	}
\startdata
GG Tau  &1.5$\times$10$^{-10}$  & 2.8$\times$10$^{-8}$  & 3.6$\pm$2.0$\times$10$^{-3}$ \tablenotemark{c} & 1.3$\pm$0.1$\times$10$^{-3}$ \tablenotemark{d} & 0.116$\pm$0.052 \tablenotemark{e}   \\
LkCa 15 &3.7$\times$10$^{-11}$  & 7.0$\times$10$^{-9}$  & 8.6$\pm$4.3$\times$10$^{-3}$ \tablenotemark{f} & \nodata                                        & 0.18\tablenotemark{g} , 0.024\tablenotemark{h}, 0.033\tablenotemark{i} \\
DoAr 21 &4.4$\times$10$^{-10}$  & 8.1$\times$10$^{-8}$  & \nodata	                                 & \nodata                                        & $<$0.0012 \tablenotemark{j}\\
TW Hya  &3.5$\times$10$^{-12}$  & 6.4$\times$10$^{-10}$ & \nodata                                        & 3.3 $\times$ 10$^{-5}$ \tablenotemark{k}       & 6.6 $\times$ 10$^{-5}$ \tablenotemark{k} \\
\enddata
\tablenotetext{a}{ Calculated using equation (1)}
\tablenotetext{b}{ Calculated using equation (2)}
\tablenotetext{c}{ 17 and 28 $\mu$m line emission at $\sim$ 110 K from Thi et al.\ (1999)}
\tablenotetext{d}{ $^{13}$CO line emission at $\sim$ 15 K from Dutrey et al.\ (1994)}
\tablenotetext{e}{ 2.6 mm continuum emission at $\sim$ 35 K from Guilloteau et al.\ (1999)}
\tablenotetext{f}{ Thi et al.\ (2001b)}
\tablenotetext{g}{ Model dependent fit to SED from Chiang et al.\ (2000)}
\tablenotetext{h}{ van Zadelhoff (2002)}
\tablenotetext{i}{ Duvert et al.\ (2000)}
\tablenotetext{j}{ 1.3 mm continuum emission 3-$\sigma$ upper limit from Andr\'{e} et al.\ (1990)}
\tablenotetext{k}{ Zuckerman et al.\ (1995)}
\end{deluxetable}

\begin{deluxetable}{lccc}
\tabletypesize{\scriptsize}
\tablecolumns{4}
\tablenum{3}
\tablewidth{0pt}
\tablecaption{$f$ values}
\label{tbl-3}
\tablehead{
		  \colhead{Star}
		& \colhead{warm H$_2$\tablenotemark{a}}
		& \colhead{CO\tablenotemark{b}}
		& \colhead{(sub)mm\tablenotemark{c}} \\
	}
\startdata
GG Tau      & 10$^7$   &   10$^7$   &  10$^9$      \\
LkCa 15     & 10$^8$   &   \nodata  &  10$^{9-10}$ \\
DoAr 21     & \nodata  &   \nodata  &  $\le$10$^7$      \\
TW Hya      & \nodata  &   10$^7$   &  10$^7$      \\

\enddata
\tablenotetext{a}{ Found as a ratio of columns 4 and 2 from Table 2}
\tablenotetext{b}{ Found as a ratio of columns 5 and 2 from Table 2}
\tablenotetext{c}{ Found as a ratio of columns 6 and 2 from Table 2}
\end{deluxetable}

\begin{deluxetable}{lccc}
\tabletypesize{\tiny}
\tablecolumns{4}
\tablenum{4}
\footnotesize
\tablewidth{0pt}
\tablecaption{X-ray Model Parameters and Predicted Line Fluxes}
\label{tbl-4}
\tablehead{
		  \colhead{Star}
		& \colhead{Observed F$_{v=1\rightarrow0~S(1)}$}
		& \colhead{log L$_x$}
		& \colhead{Predicted X-ray Stimulated} \\
		  \colhead{}
		& \colhead{}
		& \colhead{}
		& \colhead{F$_{v=1\rightarrow0~S(1)}$} \\
		  \colhead{}
		& \colhead{}
		& \colhead{}
		& \colhead{(ergs s$^{-1}$ cm$^{-2}$)} \\
	}
\startdata
TW Hya		& 1.0$\pm$0.1$\times$10$^{-15}$ & 30.3   & 6.7~$\times$~10$^{-16}$$\leq$~F$_{v=1\rightarrow0~S(1)}$~$\leq$~5.0~$\times$~10$^{-15}$ \\
GG Tau A	& 6.9$\pm$0.5$\times$10$^{-15}$ & 29.4   & 5.0~$\times$~10$^{-16}$$\leq$~F$_{v=1\rightarrow0~S(1)}$~$\leq$~5.0~$\times$~10$^{-15}$ \\
DoAr 21		& 1.5$\pm$0.9$\times$10$^{-14}$ & 31.2   & 2.4~$\times$~10$^{-17}$$\leq$~F$_{v=1\rightarrow0~S(1)}$~$\leq$~2.4~$\times$~10$^{-14}$ \\
LkCa 15		& 1.7$\pm$0.2$\times$10$^{-15}$ & 29.4\tablenotemark{a}   & 5.0~$\times$~10$^{-16}$$\leq$~F$_{v=1\rightarrow0~S(1)}$~$\leq$~5.0~$\times$~10$^{-15}$ \\
\enddata
\tablenotetext{a}{ Assumed.}
\end{deluxetable}

\begin{deluxetable}{llcccll}
\tabletypesize{\tiny}
\tablecolumns{7}
\tablenum{5}
\footnotesize
\tablewidth{0pt}
\tablecaption{Ultraviolet Excesses for H$_2$ Detections}
\label{tbl-5}
\tablehead{
		  \colhead{Star}
		& \colhead{A$_{\rm v}$}
		& \colhead{[$U-V$]$_{observed}$}
		& \colhead{[$U-V$]$_{dereddened}$}
		& \colhead{[$U-V$]$_{excess}$}
		& \colhead{}
		& \colhead{References for} \\
		  \colhead{}
		& \colhead{(mag)}
		& \colhead{}
		& \colhead{}
		& \colhead{}
		& \colhead{}
		& \colhead{A$_{\rm v}$, $U$, $V$,} \\
		  \colhead{}
		& \colhead{}
		& \colhead{}
		& \colhead{}
		& \colhead{}
		& \colhead{}
		& \colhead{Inclination Angle} \\
	}
\startdata

TW Hya      &   0.1      & 0.86  & 0.73   &  $-$1.79  & $>$85\degr & Herbst et al.\ (1994) \\
            &   0.25     & 0.86  & 0.81   &  $-$1.71  &  & Herbst et al.\ (1994) \\
GG Tau Ab   &   3.02     & 2.73  & 1.13   &  $-$1.54  & 53\degr \tablenotemark{a} & Ghez et al.\ (1997), White et al.\ (1999), Silber et al.\ (2000) \\
            &   3.38     & 2.73  & 0.94   &  $-$1.73  &  & Ghez et al.\ (1997), White et al.\ (1999) \\
GG Tau Aa   &   0.46     & 1.42  & 1.18   &  $-$1.34  & 53\degr \tablenotemark{a} & Ghez et al.\ (1997), White et al.\ (1999), Silber et al.\ (2000) \\
            &   0.98     & 1.42  & 0.90   &  $-$1.62  &  &Ghez et al.\ (1997), White et al.\ (1999) \\
DoAr 21     &   6.2      & 3.82  & 0.53   &  $-$1.4   & \nodata  &Herbst et al.\ (1994) \\
LkCa 15     &   0.64     & 1.75  & 1.41   &  $-$0.77  & 34\degr $\pm$ 14 & Bouvier \& Appenzeller (1992) \\
            &   0.64     & 2.21  & 1.87   &  $-$0.31  &  & Hamann \& Persson (1992) \\

\enddata
\tablenotetext{a}{ Inclination angle for the circumbinary disk surrounding both GG Tau Aa and Ab.}
\end{deluxetable}

\acknowledgments{Special thanks to Tracy L. Huard for his assistance with data reduction techniques and
discussions of color-color diagrams.  Also, thanks to Phil Maloney, David Hollenbach, Didier Saumon, and Nuria
Calvet for insightful discussions.  Additional thanks to N. Calvet and Andrew Walsh for providing us with the
{\it L} band photometry and G. Herczeg for providing us with {\it FUSE} continuum and line fluxes
for TW Hya prior to publication.  We also would like to thank KPNO staff members Hillary
Mathis, Doug Williams, and Ken Hinkle for making our observing runs both productive and enjoyable.  Thanks to
Ginny Nickles for her assistance in making some of the observations reported herein.  This work is supported
by NASA grant NAG5-8295 to Vanderbilt University.}

\subsection*{Figure Captions}
\figcaption[fig1.ps]{Full spectra for target stars and calibration sources.  The spectra are not shifted in
wavelength or corrected for telluric or photospheric absorptions.  Due to instrument problems during our
Taurus-Auriga observing run, a few spectra are shifted in wavelength as evidenced by misaligned telluric
absorption features.  However, a wavelength calibration was performed for each night and in some case for
individual observations when deemed necessary to correct this systemic error.  Note an uncorrected bad pixel
in the spectra of S1 near 2.1176 $\mu$m.
}
\label{fig1}

\figcaption[fig2.ps]{Spectra containing the emission lines associated with the v = 1$\rightarrow$0 S(1)
transition at 2.12183 $\mu$m.  Shifted to the systemic velocity for each source, the emission is nearly
centered at the rest velocities of the sources, to within errors.  
}
\label{fig2}

\figcaption[fig3.ps]{{\it JHK} color-color diagram of the four TTS detected in emission at 2.12183 $\mu$m.  Only
the interstellar reddening vectors for K1V and K7V stars are plotted.  Since the difference between the K5V,
K7V, and M0V vectors are barely the thickness of the plotted line, vectors corresponding to the K5V and M0V
spectral types are not plotted and can be estimated from the K7V line.  According to these vectors, only
LkCa~15 shows clear evidence of a shift due to circumstellar reddening, while GG~Tau~Aa and GG~Tau~Ab show
only a marginal shift and remain within errors of the reddening vectors.  The crosses correspond to
$\pm$ 1-$\sigma$ errors in photometry. 
}

\label{fig3}

\figcaption[fig4.ps]{{\it JHKL} color-color diagram of the four TTS detected in emission at 2.12183 $\mu$m.
Only the interstellar reddening vectors for K1V and K7V stars are plotted.  All sources, with the exception of
DoAr~21 show clear evidence of circumstellar material with substantial shifts from the plotted interstellar
reddening vectors.  DoAr~21 is shifted marginally but its position on the diagram remains, within errors,
consistent with interstellar reddening along the K1V reddening vector.  The crosses correspond to
$\pm$ 1-$\sigma$ errors in photometry.  
}

\clearpage

\begin{figure}
\plotone{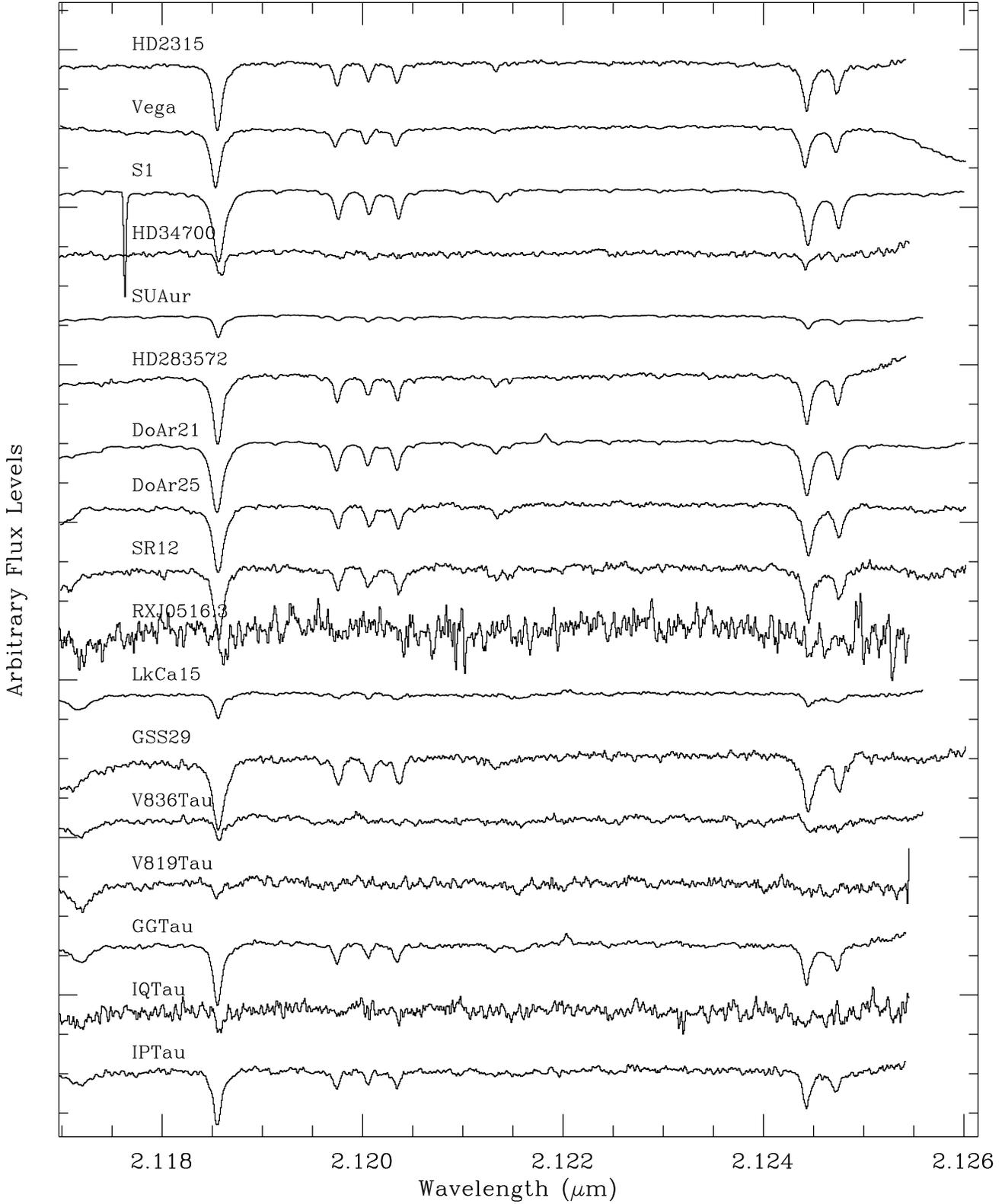}
\figurenum{1}
\caption{Source spectra}
\end{figure}

\begin{figure}
\plotone{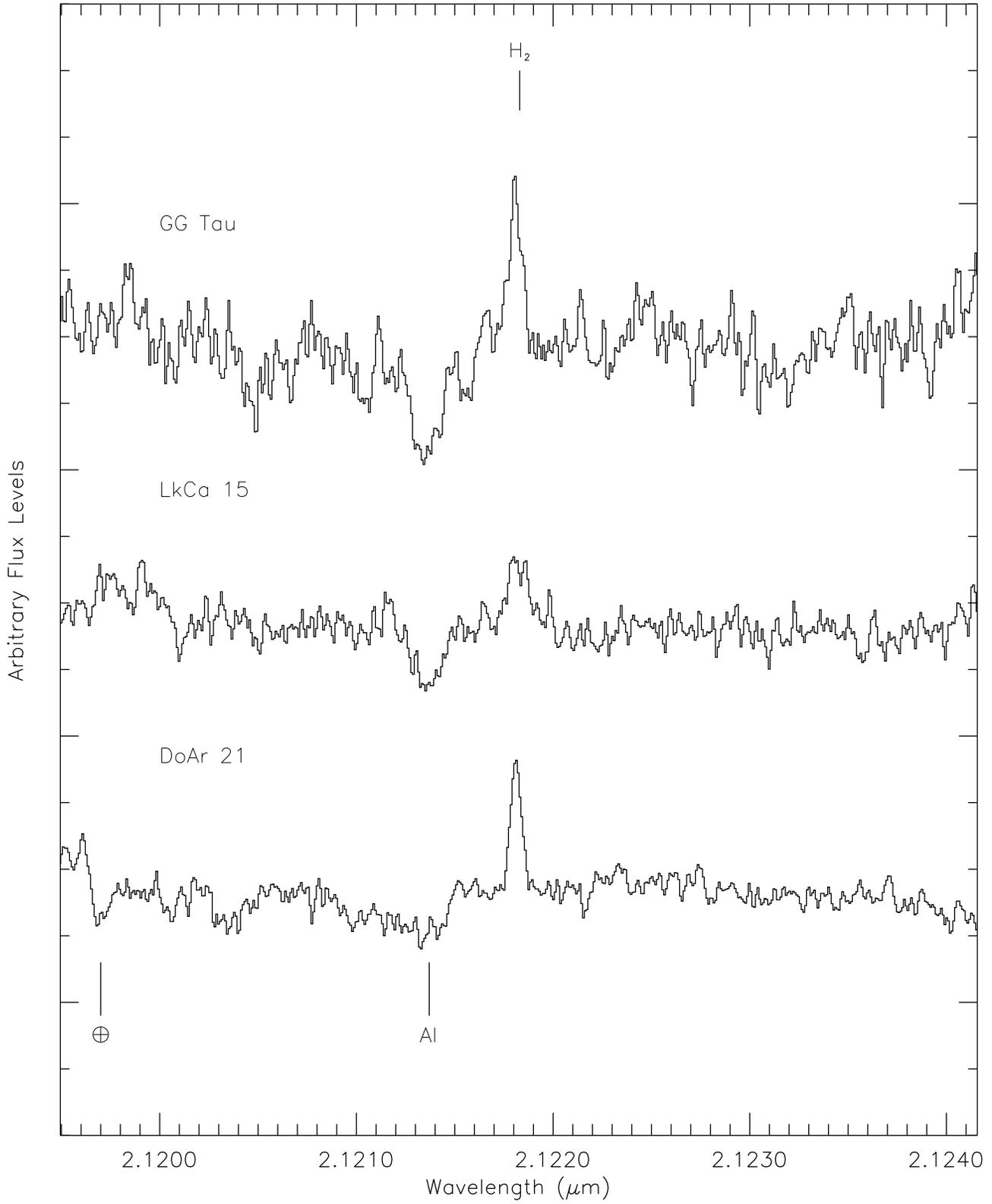}
\figurenum{2}
\caption{Emission spectra}
\end{figure}

\clearpage

\begin{figure}
\plotone{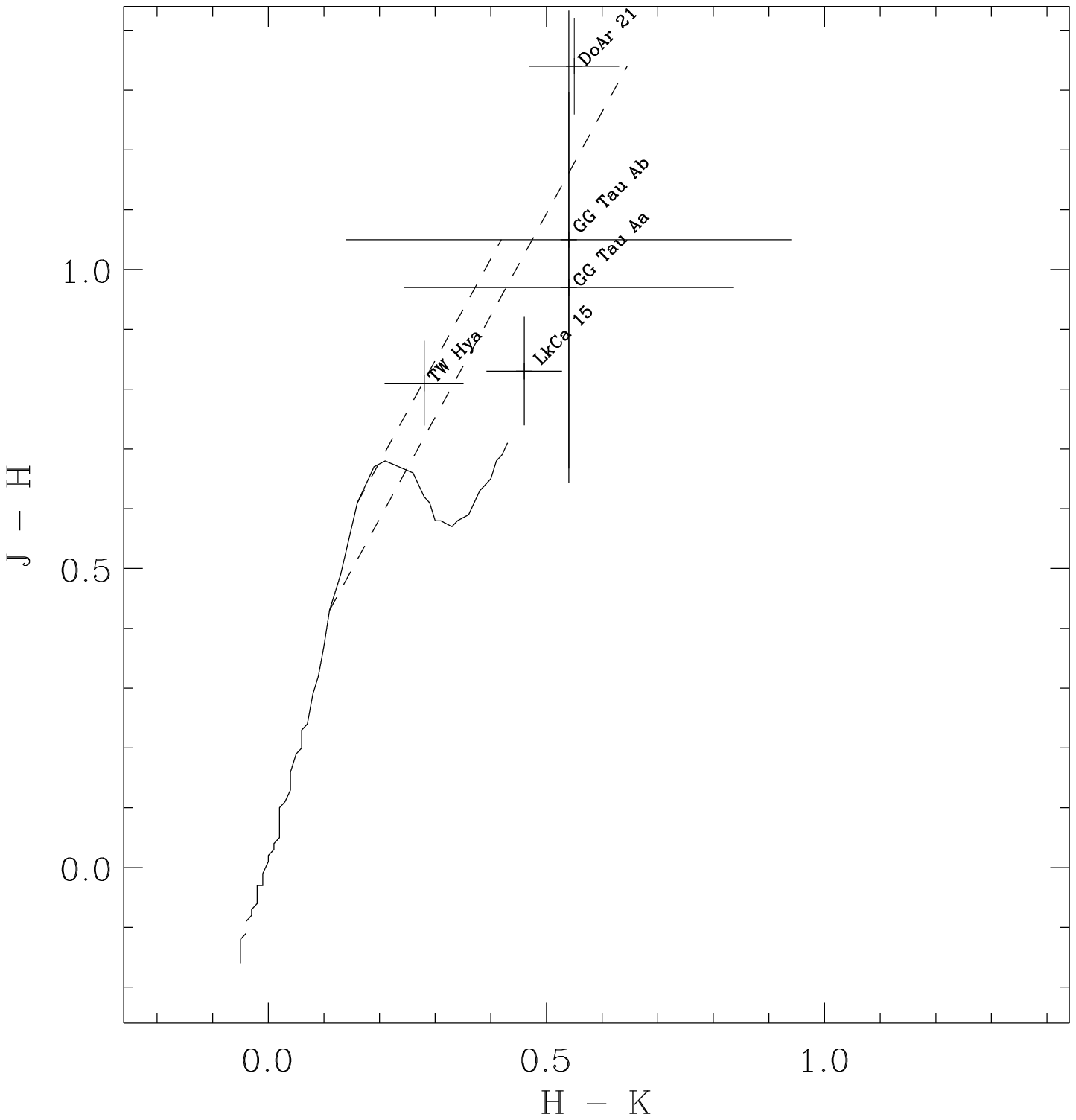}
\figurenum{3}
\caption{{\it JHK} Color-Color Diagram}
\end{figure}

\clearpage

\begin{figure}
\plotone{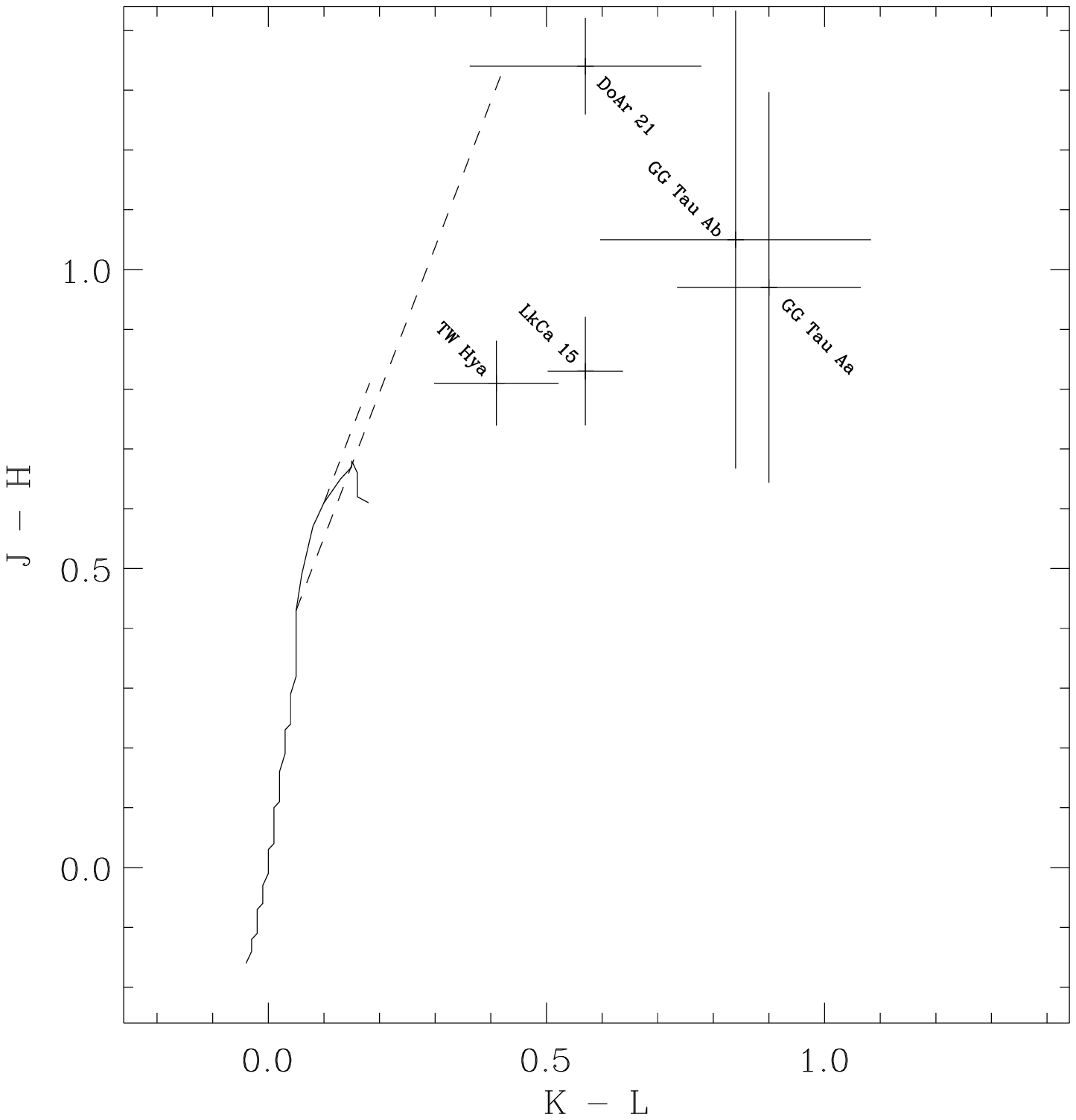}
\figurenum{4}
\caption{{\it JHKL} Color-Color Diagram}
\end{figure}

\end{document}